\documentclass[reprint,amssymb,superscriptaddress,aps,prf,longbibliography]{revtex4-1}

\usepackage{graphicx}

\usepackage{bm}
\usepackage{float} 

\usepackage{xcolor}
\usepackage{graphics}

\usepackage{color}
\usepackage{amsmath}
\usepackage{mathtools}
\usepackage{eucal}
\usepackage{lineno}

\usepackage{changes}
\graphicspath{ {images/} } 

\begin{document}
\title{Local evaporation flux of deformed liquid drops}

\author{Pan Jia}
\affiliation{School of Science, Harbin Institute of Technology, 518055 Shenzhen, China}
\affiliation{École Polytechnique Fédérale de Lausanne, 1015 Lausanne, Switzerland}

\author{Mo Zhou}
\affiliation{School of Physics, Northwest University, 710127 Xi'an, China}

\author{Haiping Yu}
\email{yuhp@nwu.edu.cn}
\affiliation{School of Physics, Northwest University, 710127 Xi'an, China}

\author{Cunjing Lv}
\affiliation{Department of Engineering Mechanics, Tsinghua University, 100084 Beijing, China}

\author{Guangyin Jing} 
\email{jing@nwu.edu.cn}
\affiliation{School of Physics, Northwest University, 710127 Xi'an, China}

\date{\today}

\begin{abstract}
Escaping of the liquid molecules from their liquid bulk into the vapour phase at the vapour-liquid interface is controlled by the vapour diffusion process, which nevertheless hardly  senses the macroscopic shape of this interface. Here, deformed sessile drops due to gravity and surface tension with various interfacial profiles are realised by tilting flat substrates. The symmetry broken of the sessile drop geometry leads to a different evaporation behavior compared to a drop with a symmetric cap on a horizontal substrate.  Rather than the vapour-diffusion mechanism, heat-diffusion regime is defined here to calculate the local evaporation flux along the deformed drop interface. A local heat resistance, characterised by the liquid layer thickness perpendicular to the substrate, is proposed to relate the local evaporation flux. We find that the drops with and without deformation evaporate with a minimum flux at the drop apex, while up to a maximum one  with a significantly larger but finite value at the contact line. Counterintuitively, the deviation from the symmetric shape due to the deformation on a slope, surprisingly enhances the total evaporation rate; and the smaller contact angle, the more significant enhancement. Larger tilt quickens the overall evaporation process and induces a more heterogeneous distribution of evaporative flux under gravity. Interestingly, with this concept of heat flux, an intrinsic heat resistance is conceivable around the contact line, which naturally removes the singularity of the evaporation flux showing in the vapour-diffusion model. The detailed non-uniform evaporation flux suggests ways to control the self-assembly, microstructures of deposit with engineering applications particularly in three dimensional printing where drying on slopes is inevitable.  
\end{abstract}

\maketitle

\section{\label{Introduction}Introduction}
Evaporation happens in nature where liquid is present, and in industry where materials form from drying of suspension solutions.
The escaping liquid molecules into the surrounding media by evaporating processes marks mass and heat transfer and pattern formation in the sense of flow-driven self-assembly of particles enclosed, therefore offers a vast number of practical applications.  Paintings, printings, coatings and films drying from solutions are heavily dependent on the controlled evaporation processes. 
The simplest scenario is that a sessile drop with a spherical cap loses its volume by evaporation until its fate. 
The evaporation flux is essentially one of the most important driven factors controlling the drop fate and therefore the final structures with special functionalities from drying solutions. 
Although this simple phenomenon happens everywhere, the hydrodynamics is hard to be taken for granted. 
Very early, it was explained by the essentials of the diffusion process \cite{MaxellBook1877, PhysRev1918Langmuir} from the observation by Morse \cite{AmAcaArtSci1910Morse} for the evaporation of a perfect spherical drop or a hemisphere, with the analogy of electrostatic potential following the similar Laplace equation. 
Using this principle, Picknett and Bexon \cite{Picknett1977evaporation} obtained an analytical form of the evaporation rate, which was later discussed analytically as well for a drop sitting on a flat substrate \cite{Langmuir1995Bourges}.  
Arising from the physical fundamentals and practical applications, extensive studies have been conducted on the evaporation \cite{Bruning2020,Li16756,Yu2013, Yu2012} and evaporation-driven flow inside the drop of complex fluids towards a variety of deposit structures  \cite {RMP1997EvolutionLiquidFilm, Tan2019, Larson2014transport, Giorgiutti2018drying, Brutin2018recent,Jaiswal2018,Gao2018}. 

With the simplification of a spherical cap for a drop evaporation on the flat substrate, the vapour concentration and its gradient for evaporation flux are possibly achieved by solving the Laplace equation. Nevertheless it is still difficult to have a precise analytical formula with an accurate form for the local evaporation flux. 
Deegan and other colleagues \cite{Deegan1997capillary, PRE2000Deegan, Popov2005PRE} treated a flat drop and conducted the diffusion-limited analysis based on Lebedev's mathematical solution for the Laplace equation \cite{lebedev1965special}. Furthermore a practical formula of evaporation was  nicely derived by this diffusion-limited mechanism, and was later confirmed by using the finite element method in the seminal reports by Hu and Larson \cite{HuJPCB2002, Hu2005Langmuir}. 
This diffusion-limited regime successfully predicts the local evaporation flux along the profile, however, with a singularity at the contact line \cite{PRE2000Deegan, HuJPCB2002}, which has been used widely and successfully in the later research towards the  explanation of the effect of non-uniform evaporative flux on the induced liquid flows and various deposits \cite{berteloot2008evaporation, jing2010drying, cira2015vapour, stauber2015lifetimes, carrier2016evaporation}. 
In this model, a \textit{random-walk} picture was proposed  that the diffusion is enhanced close to the contact line, due to the fact that the liquid molecules leave the surface with less hitting back onto the liquid surface because of the dry solid substrate outside of the drop. 
\textcolor{black}{Rednikov and Colinet developed the new model by proposing the Kelvin effect to successfully solve the sigularity problem of evaporation of volatile liquids in \it ambient air \rm at the contact line, which is quite nice and practical even for drops on total wetting surface with nanoscopic thin liquid layers \cite{Rednikov2019,Rednikov2021}}
It was realised that the vapour-diffusion model is so far the better explanation with the {`}random-walk' hypothesis for the vapour molecule,  however this model finds its difficulty when extending the cases of  drop evaporation on hydrophobic or superhydrophobic substrates \cite{xie2016evaporation}. 
 
Alternatively, Ajaev {\it et al.} calculated the evaporation flux by considering a heat balance between the vaporisation latent heat at the drop free surface and the diffused heat across the vertical liquid film, herein named as Ajaev model \cite{Ajaev2001steady, Ajaev2005spreading}. This energy balance between the phase transition and heat transport across both the substrate and drop layer\cite{schrage1953theoretical}, provides more generality to determine the evaporation flux at the interface. 
This so called one-sided model was later used to calculate the evaporation flux, therefore to investigate the flow, Marangoni instability, and deposit formation \cite{Sultan_2005JFM, Murisic2011evaporation, Espin2014sagging}.
Importantly, the one-side model avoids solving the vapour diffusion equation outside the drop, but instead solves the profile evolution governed by a higher order differential equation. 
Rather than the spherical cap approximation for the drop shape in the {\textcolor{black}{vapour-diffusion model}}, the Ajaev model can deal with the contribution of the adsorbed microscopic film at the contact line, which also naturally removes the singularity of the evaporation flux encountered in the {\textcolor{black}{vapour-diffusion model}}.
Homsy and Amini extended this method further to the evaporation drops with a moving contact line \cite{HomsyPRF1}, and on the substrates with periodic or quasi-periodic structures\cite{HomsyPRF1}, but still on horizontally orientated plates in order to keep the convenient boundary conditions by the symmetry nature. 

Although these two approaches are widely used independently to calculate the local evaporation flux along the drop free interface, the drop on curved or tilted substrates is less studied. As reviewed on the drop deformation due to the gravity and pinning process \cite{Thampi2020beyond}, it is agreed that the surprisingly less attention was attracted for either the drops evaporation on non-horizontal substrates or deformed drops.  Practically, the drop on non-horizontal plates is critically important, particularly in the applications for the 3D printing technology where fresh solution jet is always evaporating on its own deposit with highly curved facets \cite{Kong20163d, Nayak2019review, Dhar2020}. Espin and Kumar built a model system of drop evaporation on a slope and used the one-side model to estimate the evaporation flux then combine the lubrication approximation with the convection-diffusion of solute inside the drop towards the deposit patterns \cite{Espin2014sagging}. Recently Timm {\it et al.} solved the Laplace equation for vapour diffusion to calculate the evaporation flux. For this purpose, the perturbation approximation combined the numerical method were proposed to demonstrate the slope-dependent total evaporation rate and an enhancement of evaporation on slope was observed\cite{Timm2019evaporation}. However, on the contrary, Kim {\it et al.} measured the lifetime of drop drying on tilted substrate, and observed the longest total evaporation time on the perfectly vertical substrate, which was ascribed to the shape effect and pinning-depinning process by wetting hysteresis \cite{kim2017evaporation}. We note that, very recently, Charitatos \it {et. al} \rm  systematically analysed the drop fate on inclined substrates by numerical method towards the profile evolution equation, and pointed out the roughness effect on these contrary results of the total drying time. The detailed textures on substrates for example are very important for drying dynamics of suspension drops in practical applications\cite{Charitatos2021}.

Here we employ the one-sided model which can generally treat evaporating drops with the irregular shapes, i.e. deformed drops on flat plates with different tilt angles. 
Dominated by the surface tension at the drop free interface, regulations of the drop profiles are numerically determined by solving the Young-Laplace equation. Thus a second-order differential equation rather than the fourth-order evolution equation of the drop profile is numerically solved here. 
The calculated drop shaides the detailed thickness of the liquid film for heat diffusion, thereafter determines the local evaporation flux at the vapour-liquid interface. 
Vapour diffusion limited regime above the drop assumes the transport of the vapour molecules in the vapour phase, but it is not expected to be sensitive to the local shape of the vapour-liquid interface, which is always sensed as microscopic `flat' for molecules. 
However, the deformed interface changes the local liquid film thickness then the heat resistance, consequently varies the local evaporation flux. 
Different tilt angles are imposed to the substrates, leading to the enhancement of the total evaporation rate at larger tilt angles, which is counterintuitive in vapour diffusion regime. 
The singularity of the evaporation flux at the contact line is removed with a clear physical interpretation due to the film thinning, which is quantitatively compared to the reports in literatures.

\section{\label{Modelling}Modelling}
\begin{figure*}[htb!]
\includegraphics[width = 0.65\textwidth]{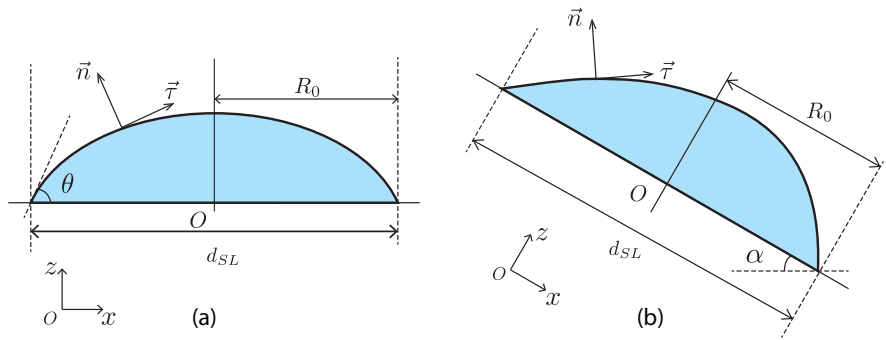}
\caption{Schematics of 2D liquid drops. (a) On a horizontal substrate. (b) On a tilted substrate. In both panels, $d_{SL}$ is the contact length and $R_0$ is the radius of the drop base. $\vec{n}$ and $\vec{\tau}$ are the unit vectors in tangential and normal directions, along the vapour-liquid interface respectively. $\theta$ is the equilibrium contact angle when the substrate is horizontal, and $\alpha$ is the tilt angle of the substrate respect to the horizontal direction. It should be noted that $x$ and $z$ axis are always along and perpendicular to the substrate, respectively, and that the contact line is pined when $\alpha$ varies.}
\label{DropSchematics}
\end{figure*}
The basic idea here is that the evaporation is dominantly controlled by the thermal diffusion across the drop from a heat reservoir of the supporting substrate, compared to the convective heat transfer along the drop height direction. Considering the general case, we take a water drop with the base radius $R_0 \sim 1~$mm, height $h\sim$ 0.5 mm, density $\rho$, vapour density of $\rho_v$ and diffusion coefficient of $D_m\sim 10^{-5}$ m$^2$/s in vapour phase, heat diffusivity $D_T \sim 10^{-7} $ m$^2$/s, and characteristic velocity $U \sim 5 ~\mu$m$/$s of flow inside the drop, and it allows us to estimate several time scales here \cite{Larson2014transport}. With these typical values, the typical evaporation time is then $t_f\sim \rho R_0 h/(\rho_{v}D_m) \sim 200 ~ $s, much longer than the heat transfer time $h^2/D_T \sim 1$ s and the vapour diffusion time $R_0h/D_m \sim 1 ~  $s.  The ratio of heat convection to the diffusion is $Uh/D_T \sim  0.01$, showing the domination of the heat diffusion process over the convective contribution. We also note that the estimated time scale of heat diffuse is  much longer than that of the phase change, which  is characterised by the thermal velocity of the molecule at the interface; and this confirms our basic idea that heat diffusion dominates the evaporation. By the heat balance,  the energy loss due to vaporisation needs to be replenished by the heat transfer from drop interior and thus induces the cooling at the vapour-liquid interface. Therefore, at first, the energy conservation relates the evaporation flux to the heat diffusion across the liquid film with the local thickness. It turns out that the vapour diffusion governs the distribution of the vapour density and then the evaporation process of the drop, whereas the heat balance sets the vertical temperature distribution inside the drop. The one-sided model is employed here, with the convenience, to determine this evaporative mass flux from the energy conservation. 

Following this idea, we consider here a two-dimensional volatile liquid drop pinned on a flat substrate, with orientations tuned by the tilt angles and with a constant uniform temperature $\tilde T_0$, evaporating in its pure vapour environment  (Fig.~\ref{DropSchematics}). We introduce the capillary number $ C={\mu \tilde U}/{\gamma}$, with $\mu$ the dynamical viscosity, and $\gamma$ the surface tension at the equilibrium saturation temperature $\tilde T_s$. 
The characteristic velocity reads as $\tilde U={k \tilde T_s}/{(\tilde \rho \mathcal{L} R_0)}$, followed the definition in the Ajaev model \cite{Ajaev2005spreading}, with $k$ and $\mathcal{L}$ being respectively the thermal conductivity and evaporation latent heat per unit mass of the liquid, and the contact length $d_{SL}=2R_0$ (Fig.~\ref{DropSchematics}). Normally, the evaporation drops run in the low limit of capillary number $C$. Under the lubrication approximation, the coordinates $( \tilde x,\tilde z)$, velocity field $(\tilde u,\tilde w)$, temperature $ \tilde T$, pressure $ \tilde p$ and evaporation flux $\tilde J$, have the dimensionless forms as
\begin{eqnarray}
x= \frac{\tilde x}{R_0}\,,\qquad z=\frac{\tilde z}{R_0 C^{1/3}}\,, \qquad u=\frac{\tilde u}{\tilde U}\,, \qquad w=\frac{\tilde w}{\tilde U C^{1/3}}\,,
\nonumber
\\
p=\frac{\tilde p}{C^{1/3} \gamma / R_0}\,, \qquad  T=\frac{\tilde T-\tilde T_s}{C^{2/3}\tilde T_s}\,, \qquad J=\frac{\tilde J}{C^{1/3}\tilde \rho \tilde U}\,.
\nonumber
\end{eqnarray}
%
Considering that the thermal diffusion is the dominant effect, we thus have the governing equations at the leading order as follows
\begin{eqnarray}
\frac{\partial u}{\partial x}+\frac{\partial w}{\partial z} & = &0\,,\label{MassConcervation}\\
-\frac{\partial p}{\partial x} +  \frac{\partial^2 u}{\partial z^2} + B_x & = &0\,, \label{NSx1}\\
-\frac{\partial p}{\partial z} - B_z & = &0\,,
\label{NSz1}\\
\frac{\partial^2 T}{\partial z^2} & =& 0,
\label{QeEnergy}
\end{eqnarray}
where $B_x=\tilde \rho g \sin \alpha R_0^2/ \left( C^{1/3}\gamma \right)$ and $B_z={\tilde \rho g \cos \alpha R_0^2}/{\gamma}$, with $g$ the gravity acceleration and { $\alpha$ the tilt angle of the substrate with respect to the horizontal direction (Fig. \ref{DropSchematics}b).} The inertia and thermal convection in the energy conservation equation (Eq. \ref{QeEnergy}) are acceptably neglected here, as well in the {\textcolor{black}{vapour-diffusion model}} and heat-diffusion model. No-slip condition at the liquid-solid interface, and the temperature continuity read as
\begin{eqnarray}
u=w  = 0 \quad &{\rm for}& \quad z=0 \,,
\label{BCSubV}\\
T=T_0 \quad &{\rm for}& \quad z=0 \,.
\label{BCSubT}
\end{eqnarray}
The local height of the vapour-liquid interface (i.e., the drop height) is denoted as $h(x)$, where the normal stress is $ \vec{n}\cdot \tilde \Sigma\cdot\vec{n} =-\tilde p$, with $\tilde \Sigma$ the stress tensor in the liquid side at the interface and $\vec{n}$ the outward normal unit vector. Here the disjoining pressure is neglected, but will be qualitatively discussed later for the singularity problem of the evaporation flux. $p_v$ is the rescaled vapour pressure, assumed to be constant over the whole interface. The stress balances in the normal and tangential directions at the vapour-liquid interface are written as
\begin{eqnarray}
p - p_v = \kappa(x) - \frac{\epsilon}{h^3}\,,
\label{NNormalStress1}
\end{eqnarray}
\begin{equation}
\vec{n}\cdot\Sigma\cdot\vec{\tau} = \frac{\partial u}{\partial z} =0\,.
\label{NShearStress2}
\end{equation}
where $\kappa (x)=- \partial^2 h/ \partial x^2/(1+\partial h/ \partial x)^{3/2}$ the curvature of the interface; \textcolor{black}{$\epsilon / h^3$ represents the contribution from the disjoining pressure and $\epsilon$ is the rescaled Hamaker constant, estimated at the order of $10^{-3}$ in the limit of small $C$ \cite{Ajaev2005spreading, Ajaev2001steady}.} For the energy condition, the heat flux transported through the droplet equals to the latent heat of the evaporating vapour

\begin{eqnarray}
\centering
J = -\frac{\partial T}{\partial z} |_{z=h}\,.
\label{LJ}
\end{eqnarray}
For the evaporation of the liquid in its pure vapour environment, the evaporation flux at the interface is generally controlled by the pressure difference $\tilde p_v^e - \tilde p_v$, using commonly the Hertz-Knudsen relation \cite{Persad2016}. $\tilde p_v^e$ is the thermodynamical equilibrium pressure, which is determined by the interfacial pressure and temperature. According to Schrage-Ajaev-Homsy's calculations\cite{schrage1953theoretical, Ajaev2001steady}, $\tilde J$ is specified as
\begin{eqnarray}
\tilde J=\tilde \rho_v \left(\frac{\tilde R \tilde T_i}{2 \pi}\right)^{\frac{1}{2}}\left(\frac{\tilde p_v^e}{\tilde p_v}-1\right)\,,
\label{EFlux}
\end{eqnarray}
where $\tilde \rho_v$ is the vapour density, $\tilde T_i$ is the interfacial temperature and $\tilde R$ is the specific gas constant. Eq. \ref{EFlux} is further expressed in a dimensionless way
\begin{eqnarray}
K J=\delta \left(p-p_v\right) + T_i\,,
\label{EvaFlux1}
\end{eqnarray}
with $K=\left(\tilde \rho \tilde U \sqrt{2 \pi \tilde R \tilde T_s}\right)/\left(2 \tilde \rho_v \mathcal L C^{1/3}\right)$ and $\delta=\gamma/\left({\mathcal{L} \tilde \rho R_0 C^{1/3}}\right)$ are dimensionless  parameters originated from the Hertz-Knudsen relation (Eq. \ref{EFlux}), thus indicating respectively the kinetic effect and pressure variation effect at the interface \cite{Persad2016}.

Obviously, in order to calculate the evaporation flux, the pressures across the vapour-liquid interface and its temperature are needed. This pressure jump can be found in the normal stress balance (Eq.\ref{NNormalStress1}), where the drop profile $h(x)$ is correlated. The interfacial temperature is derived from the heat balance Eq. \ref{LJ}
\begin{equation}
\label{eq:interfacialT}
T_i = T_0-Jh\,,
\end{equation}
where we have used the temperature boundary condition at the substrate (Eq. \ref{BCSubT}). The drop profile has to be determined firstly, which asks for the velocity field inside the drop by the mass conservation relation for the evaporation-induced flows. 

The pressure $p$  in Eq. \ref{NSz1} is decoupled into two components of height dependent $-B_zz$ and $x$ dependent $p_1$ as
\begin{eqnarray}
p=-B_z z +p_1\,.
\end{eqnarray}
Integrating twice the momentum conservation (Eq. \ref{NSx1}) with respect to $x$  gives the velocity profile $u$; and $w$ is further obtained via the mass conservation (Eq. \ref{MassConcervation}) 
\begin{eqnarray}
u&=&\frac{1}{2}\left(\frac{\partial p_1}{\partial x}-B_x\right)\left(z^2-2 h z\right)\,,\label{U}\\
w&=&-\frac{1}{2}\frac{\partial^2 p_1}{\partial x^2}\left( \frac{1}{3}z^3 - h z^2\right) + \frac{1}{2}\left(\frac{\partial p_1}{\partial x}-B_x\right)\frac{\partial h}{\partial x}\, z^2\,,\nonumber\\
\label{W}
\end{eqnarray}
where boundary conditions Eqs. \ref{BCSubV} and \ref{NShearStress2} are used. The pressure at the $z=h(x)$ is
\begin{eqnarray}
\frac{\partial p_1}{\partial x} =\frac{\partial}{\partial x} \left( \kappa (x) -\frac{\epsilon}{h^3}+B_z h \right)\,.
\label{p1x}
\end{eqnarray}

As for the evaporation rate, it is determined by the interfacial temperature $T_i$ and pressure difference. Plugging further Eqs. \ref{NNormalStress1} and \ref{eq:interfacialT} into \ref{EvaFlux1} gives
\begin{eqnarray}
\textcolor{black}{J = \left(T_0 +  \delta \kappa (x) - \delta \epsilon /h^3\right)/R_h\,.}
\label{Jdis}
\end{eqnarray}
\textcolor{black}{For a drop with the typical dimension larger than the characteristic capillary length $l_c=(\gamma / \rho g)^{1/2}$,  it is reasonable to treat this drop as two parts, the macroscopic curved air-liquid interface away from the contact line and the corner region within the contact line. It is inevitable that  the disjoining pressure plays the important role to consider the microscopic film close to the contact line, which is included in the Eqs.\ref{Jdis}. However, as being calculated in the next section, the $\delta \epsilon/h^3$ is negligible when the heat diffuses across the liquid layer much thicker.  Therefore, the contribution of disjoining pressure will be negligible, compared to $T_0$ and $\kappa(x)$,  for the drop here deformed by the capillarity and gravity. Therefore, the equation of evaporation flux above is approximately simplified as}
\begin{eqnarray}
J = \left(T_0 +  \delta \kappa (x)\right)/R_h\,,
\label{JJ}
\end{eqnarray}
where  $R_h = K+h$, is defined here as the thermal resistance of across the drop thickness, and accordingly, $T_0 +  \delta \kappa (x)$ is the temperature difference. Therefore, Eq. \ref{JJ} is essentially the Fourier's law describing the heat diffusion process from the substrate to the vapour-liquid interface. Obviously, these results indicate that the flow inside the drop and the evaporation flux are directly related to the drop profile $h(x, t)$, which is also the only {unknown} in the expressions (\ref{U} - \ref{JJ}). \\

Considering the mass conservation at the interface, we have ${\partial h}/{\partial t} = - u\,{\partial h}/{\partial x} + w - J $. Combining these terms obtained above, the evolution equation reads as
\begin{eqnarray}
h_t - \frac{ \delta \kappa (x) + T_0 }{K+h}  + \frac{1}{3} \left[\left( \kappa (x) + B_z h\right)_x h^3 \right]_x + \frac{1}{3} B_x (h^3)_x=0\,.\nonumber\\
\label{Ht}
\end{eqnarray}
{Note that the subscripts $x$ and $t$ in the equation above indicate the derivatives in $x$ direction and in time, except $B_x$}. Generally, this fourth-order partial differential equation has to be solved in a numerical way, given the symmetric boundary conditions for $h_{xxx}(0)$ and $h_x(0)$ at the drop apex, as performed by Ajaev and Homsy\cite{ajaev2012interfacial, HomsyPRF1}. However, the problem in the present case loses this symmetry due to the gravity effect, which makes the numerical method for Eq. \ref{Ht} even more complicated.  Fortunately, for drops with sizes comparable to the capillary length { $l_c$, the surface tension $\gamma$ regulates }the drop shape immediately, being at a time scale shorter than that of the vapour and heat diffusion. With this fact, instead of directly solving  Eq. \ref{Ht}, we compute herein the drop profile from the Young-Laplace law. Gomba {\it et al.} (2009) and Perazzo {\it et al.} (2017) respectively found the analytic solutions for the profiles of two-dimensional droplets equilibrium with a surrounding thin film, and a finite-length precursor film on a solid substrate \cite{Gomba2009,Perazzo2017}.

When the gravity effect comes into play, the drop profile is governed by
\begin{eqnarray}
\kappa^* \gamma = \rho g z^* + \Delta p_0\,,
\label{YL}
\end{eqnarray}
where $\Delta p_0$ is a constant, $\kappa^*$ is the curvature and $z^*$ is the vertical coordinate when calculating the profiles (See {\it Appendix}). Landau and Lifshitz calculated the profile of a drop on a horizontal substrate \cite{landau1959fluid}. Tanasijczuk {\it et al.} solved the general thickness of profiles for sessile and hanging two-dimensional droplets on arbitrarily shaped substrates under gravity \cite{Tanasijczuka2010}. In 2012, Gomba {\it et al.} gave an analytical solution for partially wetting two-dimensional droplets when surface tension, gravity and disjoining pressure are included \cite{Gomba2012}. More recently, Lv and Shi solved analytically the profiles of drops on tilted substrates \cite{Lv2018PRE}; with the method presented in this paper, we are able to compute the drop profiles on tilt substrates. In the {\it Appendix}, we present the detailed information of computing profiles, including an experimental verification. 
%
%
%

With the obtained profile, we are then able to calculate the evaporation flux along the interface. Fortunately at last, this alternative numerical method for the profile equation controlled by the Young-Laplace equation modulated by the surface tension and gravity, paves the way for the local evaporation flux. 

\begin{figure*}[htb]
\includegraphics[width=0.8\textwidth]{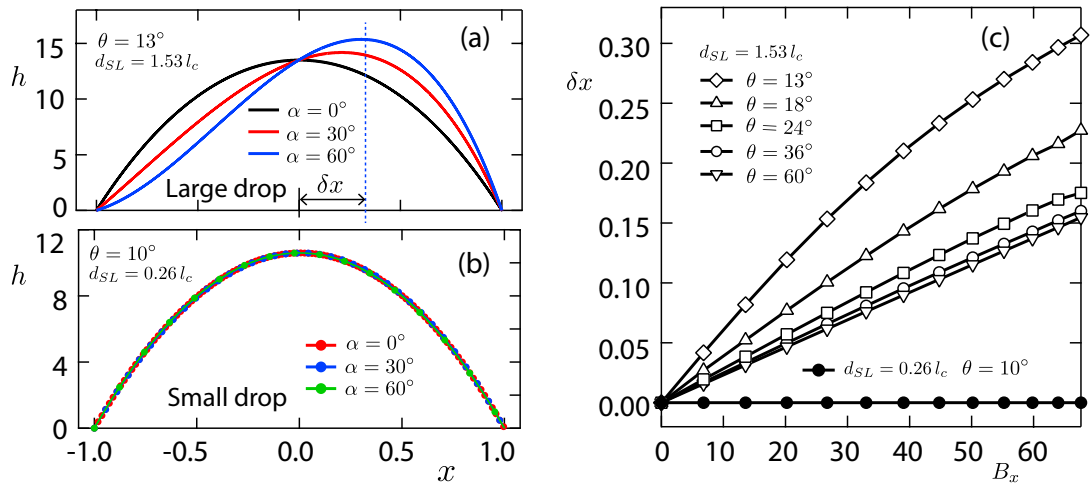}
\caption{Drop profiles. (a) Drop with contact diameter $d_{SL} =1.53\,l_c$ on partial wetting substrate $\theta$  = 13$^\circ$, subjected to tilt angles $\alpha$  = 0$^\circ$ , 30$^\circ$ , 60$^\circ$. The deformed profile shows the thinner upper part and thicker low part at high tilt {angles}. $\delta x$ is the deviation distance in the $x$ direction for the profile apex.   (b) Small {drops} keep approximately the spherical cap, with the contact diameter $d_{SL} = 0.26\,l_c$. (c) Tilt effects on the profile apex under different tilt angles and wetting states. It should be noted that, in panels (a)  and (b), we have done a counter-clock $\alpha$ rotation for the profiles on tilted substrates, in order to have all the contact lines overlapping on the one on the horizontal substrate, and therefore to achieve a clear comparison.} 
\label{Profile}
\end{figure*}
\section{Results and discussions}\label{Results and discussions}

\subsection{Drop profile}
For large drops, we take the contact diameter larger than the capillary length, $d_{SL} = 1.53\, l_c$. Accordingly, the profiles are numerically computed for tilt angles ranging from  $\alpha=0^\circ$ to 60$^\circ$; five contact angles $\theta$  = 13$^\circ$ , 18$^\circ$ , 24$^\circ$, 36$^\circ$ , 60$^\circ$ are considered for partial wetting states.

\begin{figure*}[t]
  \includegraphics[width=0.8\textwidth]{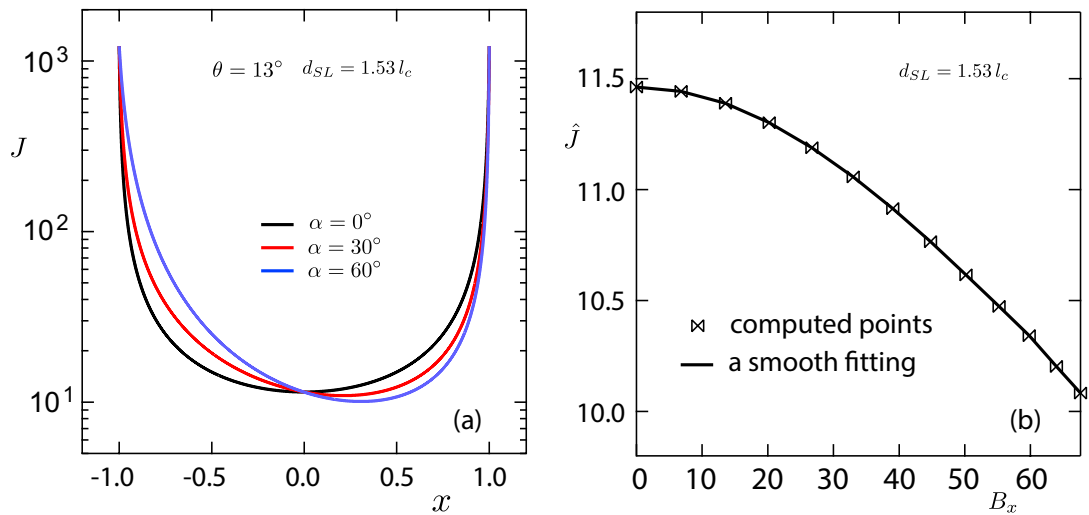}
\caption{Local evaporation flux. (a) Distribution of the evaporation rate over the vapour-liquid interface (i.e. drop profile) of a liquid drop on substrates with three different tilt angles. (b) The behaviour of the rescaled minimum evaporation rate as a function of $B_x$ for all the contact angles considered.}
\label{EvaporationRate}
\end{figure*}
In Fig. \ref{Profile}a, we present three dimensionless profiles  for the case of $\theta = 13^\circ$, obtained via the non-dimensionless definitions in the beginning of section \ref{Modelling}, and it is shown that the drop is deformed more and more with the increase of the tilt angle $\alpha$. Interestingly, all the profiles apparently meet at a same point, which separates the drop interface into two parts. In the present configuration, the left part corresponding the upper part of the drop sitting on the slope, gets thinner and thinner at larger tilt angles. Whereas, the right part displays the thickening of the drop at the lower part. Obviously, the axial symmetry with respect to the drop centre on horizontal substrate disappears. The apex, where the liquid film thickness is the highest, deviates from the centre to the right (lower) part of the drop. It is reasonable that the gravity deforms the drop, against the regulation by the surface tension. Also this is consistent with the fact that the drop size is larger than the capillary length. A characteristic parameter $\delta x$ is defined to indicate the deviation distance of apex in $x$ axis. As an example, a blue dotted line is used to show $\delta x$ for the profile of $\alpha = 60 ^{\circ}$ (Fig. \ref{Profile}a). As shown in Fig. \ref{Profile}c, $\delta x$ for all the contact angles and tilt angles are plotted.  It is indicated that $\delta x$ rises monotonically as $\alpha$ for a given contact angle due to the gravity effect, while decreases when $\theta$ increases for a given $\alpha$. As a comparison to highlight the deformation due to gravity effect, we considered also a drop with a contact diameter on the substrate smaller than the capillary length: $d_{SL} = 0.26\,l_c$. In this case, the profiles keep unchanged (Fig. \ref{Profile}b) when $\alpha$ varies, which is expected since the capillarity dominates gravity. Accordingly, $\delta x$ remains also a constant null as in Fig. \ref{Profile}c. 

By solving the Young-Laplace equation, the profiles for drops at different wetting states $\theta$ and subjected to different slopes $\alpha$ are successfully calculated. These profiles are indispensable for the calculation of the evaporation flux along the profiles. 
\subsection{Deformation effects on evaporation}
With the obtained profiles and the thermodynamic parameters, we are able to compute the local evaporation flux along the vapour-liquid interface via Eq. \ref{JJ} , where the dimensionless temperature $T_0 \approx 156$, corresponding to a normal experimental case of 25\,C$^{\circ}$. $K$ and $\delta$ are both typically small; using the involved parameters of the drop and the surrounding environment, we have accordingly the estimation of $K \approx 0.1$ and $\delta \approx 10^{-6}$. 

As shown in Fig. \ref{EvaporationRate}a, three distributions of dimensionless evaporation flux are presented. Typically, for the drop on the horizontal substrate, the minimum evaporation happens at the drop apex, where liquid layer is the thickest and thus the heat resistance peaks. The maximum one occurs at the contact line, where the liquid layer thickness and thus the heat resistance reach the minimum. These predicted behaviours of the evaporation flux are consistent with the results in reports with the vapour-diffusion model \cite{PRE2000Deegan, HuJPCB2002}. For those on the tilted substrates, similarly, the evaporation flux reaches a minimum $J_{\rm min}$ at the drop profile apex $h_{\rm max}$, whose positions in $x$ axis are illustrated in Fig. \ref{Profile}c. That's to say, the minimum evaporation flux has a corresponding dependence on the thickest liquid layer, therefore the largest heat resistance, which moves downward when the substrate tilts more and more. Considering that the contribution from the profile curvature is much smaller than that from the thermal effect: $T_0 \gg  \delta \kappa (x)$, and that $h_{\rm max} \gg K$, we have $J_{\rm min} \approx T_0/h_{\rm max}$. Thus a rescaled minimum evaporation rate is introduced $\hat J = J_{\rm min} (h_{\rm max}^{\theta}+K)/(h_{\rm max}^{13}+K)\approx J_{\rm min} h_{\rm max}^{\theta}/h_{\rm max}^{13}$, where  $h_{\rm max}^{\theta}$ is the maximum height of the drop profile at different contact angles $\theta$. As expected, a good data collapse is achieved as a function of $B_x$ for all the contact angles considered, as shown in Fig. \ref{EvaporationRate}b , where the black curve is a smooth fitting, indicating that $J_{\rm min}$ is decreasing with $B_x$. This is easy to understand since $h_{\rm max}$ is rising with $B_x$, inducing an increasing heat resistance. We have also noted that the maximum evaporation flux keeps a saturated constant, independent of the tilt angle. Possibly the microscopic thin film at the corner of the contact line dominates when the drop is subjected to the gravity. Thus, $\hat{J}$ reflects the heterogeneity of the evaporation flux distribution along the vapour-liquid interface. It indicates that the gravity effect increases the evaporation flux heterogeneity.
\begin{figure*}[tb]
  \includegraphics[width=0.8\textwidth]{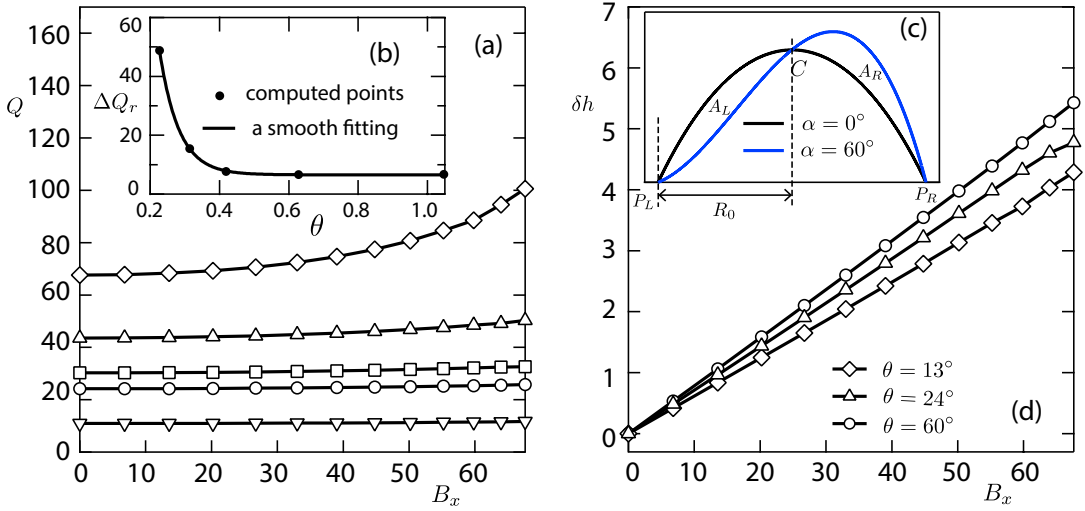}
\caption{Total evaporation rate over the drop profiles. (a) Total evaporation rate as a function of $B_x$, for the same contact length $d_{SL} = 1.53\,l_c$; five contact angles are considered, and symbol conventions are the same as in Fig. \ref{Profile} (c). (b) Relative increase of the total evaporation rate $Q$ as a function of the contact angle $\theta$. (c) Schematic of the profile height variations when the substrate is tilted. (d) Drop height variation $\delta H$ as a function of $B_x$ for three contact angles.}
\label{Q}
\end{figure*}

The total evaporation rate over the vapour-liquid interface is $Q = \int J(x) {\rm d} x$, shown in Fig. \ref{Q}a  for the dependence on $B_x$. The results indicate that $Q$ increases with $B_x$, however the relative increase drops quickly when the the substrate gets less hydrophilic (i.e. larger contact angles), shown in Fig. \ref{Q}b. Two main features are presented here. On one hand, the total evaporation rate does change due the slope effect, and drops dry faster when deformed by a slope. On the other hand, the wetting state of the drop on the substrate also plays a role in this enhancement, showing more significant enhancement if the the substrate is more hydrophilic. Thus, there is a clear enhancement of  the evaporation for the deformed drops when compared those on flat substrates, particularly for the ones with a small contact angle on the substrate, i.e., higher wettability of the substrate.

To understand such behaviour of the total evaporation rate $Q$, we present here a qualitative reasoning. As shown in Fig. \ref{Q}c, a symmetric drop profile (in black) on the horizontal substrate and a deformed drop profile (in blue)  on a tilted substrate are compared. Two profiles intersect at the point $C$, which is generally the highest point (also the central point) of the horizontal profile. This point, together with two at the contact line ($P_L$ and $P_R$), encloses two closed areas $A_L$ and $A_R$.  We further have $A_L= A_R = \Delta A$ since the liquid is incompressible. Accordingly, the average height variation, therefore the change in heat resistance,  can be estimated by $\delta h \sim \Delta A / R_0$. Recalling the evaporation flux $J \sim \frac{T_0}{h + K} $, it is obvious that $J$ increases on the left and decreases on the right. By neglecting the curvature effect, the net increase of the evaporation rate can be estimated as
\begin{equation}
\delta J \sim \frac{2 T_0 \left(h_0 + K\right)}{(h_0+K)^2-\delta h^2} - \frac{2 T_0}{h_0 + K}
\end{equation}
where $h_0$ indicates the profile on the horizontal substrate. Form this equation, we can see that, for a given $h_0$, a rising $\delta h$ would induce a rising $\delta J$, which is consistent with the fact that $\delta h$ is increasing as the tilt angle rises, as shown in Fig. \ref{Q}d. A further expansion analysis on $\delta h$ gives us $\delta J \approx 2 T_0 \delta h^2 / (h_0 + K)^3$, and thus $\delta h^2$ contributes linearly to $\delta J$. Then, the increase of the total evaporation can be estimated by $\delta Q = \int_0^{R_0} \delta J {\rm d}x$. When the contact angle increases, $h_0$ is rising, which generally dominates $\delta h$ and $K$, and thus, a larger contact angle corresponds to a smaller increase in the evaporation rate. It is reasonable also from the point view of the heat resistance represented by profile height $h$.

\subsection{Singularity and contact line}\label{CL}
A significant large evaporation flux occurs at the contact line. Due to the heat diffusion path down to a microscopic scale there, the evaporation flux sharply {increases}, but without divergence, according to the master equation (Eq. \ref{JJ}). From the time scale point of view, the thermal diffusion time scales as $\tilde h^2/D_T$, which shows a quadratic dropping to a very short time and eventually approaches the time scale of phase transition of the liquid to its vapour phase. 

Obviously, the singularity problem for evaporation at the drop edge is not present in our method. A forced decay to zero was introduced by Masoud and Felske \cite{Masoud2009POF}, whereas a non-zero thickness of a liquid film at the contact line is introduced here. Then the evaporation flux tends a finite value with the calculation of 
\begin{equation}
\label{SJJ}
 J_{\rm max} = \left(T_0 +  \delta \kappa (x)\right)/\left({K+h_{\rm micro}}\right) \sim T_0 /({K+h_{\rm micro}}) .
 \end{equation}
 Here, the $h_{\rm micro}$ is the thickness of an ultra-thin but finite film around the contact line, which has also used by de Gennes to remove the infinite viscous dissipation due to velocity divergence at the contact line \cite{1985RMP-deGennes}. Following the idea of de Gennes, the precursor film is reasonably defined here in order to remove this plausible singularity. 
 This microscopic film was also emphasised in the Ajaev model \cite{Ajaev2005spreading}, and was ascribed to the molecular interaction. 
 
 In the present work, the magnitude of the thickness is estimated as $h_{\rm micro} \sim K$, a reasonable thickness linking the  intermolecular interactions. This gives the real thickness about $2 ~ {\mathrm{\mu m}}$ , which is at the same order of magnitude ($\sim 1.4 ~ {\mathrm{\mu m}}$) defined the transition from the macroscopic drop to microscopic corner close to the contact line \cite{starov2019wetting}. If choosing a even lower dimensionless temperature $T_0$, i.e. $T_0 \sim 2$ instead of 156 in the present calculation, corresponding to a room temperature of  $22~^\circ$C, we will have this microscopic film at the thickness of $h_{\rm micro} \sim10$ nm, which is strikingly reasonable for the adsorbed film at the ambient environment.
Thanks to this ultra-thin liquid layer, the evaporation is dramatically balanced by the van der Waals interaction, therefore suppressing the divergence problem of the evaporation flux. Accordingly, we can also define an apparent heat resistance due to this microscopic film, as $R_{\rm h}\sim h_{\rm micro}$, being the counterpart of the evaporation flux singularity.
Furthermore, even the thickness of the film at contact is sharply zero, i.e. for partial wetting cases with a sharp boundary of wet and dry across the drop edge, the local evaporation flux tends to $T_0 /K$. This value is determined and independent of the tilt angle, shown in Fig. \ref{EvaporationRate}. In this figure, the dimensionless evaporation flux scales as, $J_{\rm max} \sim 10^{3}$, corresponding to the real flux of $\sim 0.03 ~ \text{kg}~\text{m}^{-2}~ \text{s}^{-1} $, which is close to the value $\sim 0.01 ~ \text{kg}~\text{m}^{-2}~ \text{s}^{-1}  $ reading from Hu-Larson's paper \cite{HuJPCB2002}.

\textcolor{black}{However, we have to admit that this intrinsic convergence of the evaporation is possibly ascribed to the microscopic film at the contact line, which has been neglected in the case where surface tension and gravity dominates for the macroscopic part of the drop. We can conveniently take the disjoining pressure into account as in Eq.  \ref{Jdis}. Since the microscopic film at the contact line comes to play the dominant role, $1/h^3$ could be much larger, and the other effects would be for sure less dominate. Thus it is interesting to further investigate the singularity just include the disjoining pressure contribution. }

\subsection{Temperature fields}
\begin{figure}[tb]
  \includegraphics[width=0.45\textwidth]{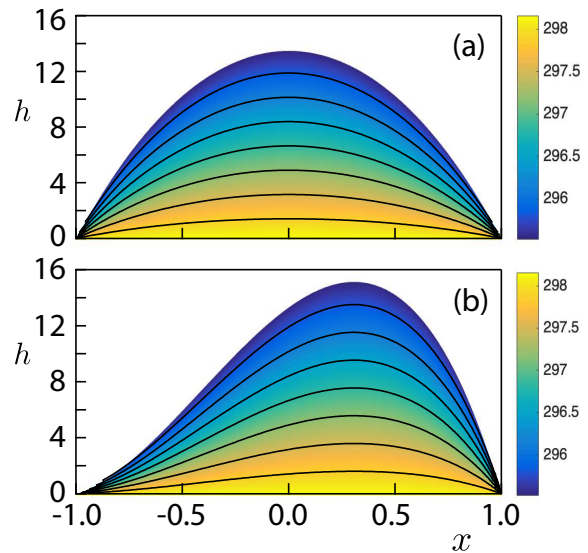}
\caption{Temperature fields inside a drop on a horizontal substrate (a), and a tilted substrate with $\alpha = 60^{\circ}$ (b); the black curves are the isothermal lines.  }
\label{Temperature}
\end{figure}

The temperature field inside the drop is obtained using the boundary conditions at the substrate (Eq. \ref{BCSubT}) and at the interface (Eq. \ref{LJ})
\begin{eqnarray}
T=T_0-J\,z\,.
\label{TT}
\end{eqnarray}
In {Fig. \ref{Temperature}}, we present the temperature fields and also the isotherms inside the drop. Obviously, the temperature decreases from the substrate to the drop profile. The temperature and isotherms are symmetric when the substrate is horizontal (Fig. \ref{Temperature}a), and predictably, they are deformed when the substrate is tilted where the gravity effect matters (Fig. \ref{Temperature}b). The isotherms are denser near the contact lines, indicating a higher heat flux there, and thus a higher local evaporation flux, which has been discussed in the previous subsection \ref{CL} from the point view of the heat resistance. It should be noted that the drop profile is not an isotherm, the temperature undergoes a minimum at the apex, and takes the maximum at the contact line, being very close to the substrate temperature $T_0$.

As discussed in the modelling section, the driving factor for the persistent evaporation is the pressure difference between the vapour at the free interface and the saturated pressure for a given working temperature. This pressure difference is transformed to a temperature difference arising between the free interface and the liquid-solid interface.
The self-cooling at the vapour-liquid interface shows that the colder free interface compared to the solid substrate with the constant temperature of $T_0$. The temperature gradient from the substrate to the vapour-liquid interface is determined by the local evaporation flux $J$, shown in Fig. \ref{Temperature}. Lastly, the Marangoni effect is expected to be present. Since the $T_0$ is in a linear way contributing to the evaporation $J$,  then the $T_s$ value just shift the loss of water by evaporation. Therefore, the $T_s$ value could change the total evaporation flux linearly, but keep the relative difference when being deformed of the drop shape on the tilted substrates. It is noted that the temperature difference across the vapour-liquid interface is sensitive to the value of $T_s$, thus hot substrate will definitely enhance the Marangoni flow across the interface, which is out of the scope of the present paper.

\section{Concluding remarks}\label{Concluding remarks}
In short,  drop evaporation on flat but tilted substrates is considered here, and the local evaporation flux is calculated. Three parameters drop size, wetting state, tilt angle are included. The local evaporation is correlated to the local thickness of the drop, or to say the details of the drop profile. Rather than solving the Laplace equation of vapour-diffusion, heat balance is used to set the evaporation flux at the vapour-liquid interface within the heat-diffusion regime. The profile of the deformed drop is conveniently calculated numerically from the Young-Laplace equation balancing the gravity and surface tension. Then the evaporation flux is explicitly calculated with the drop profile. 

The deformed drop shows the lowest evaporation flux at the apex of the vapour-liquid interface where liquid layer thickness is the largest. This flux gets larger and larger towards the contact line, with the thickness getting smaller and smaller. Interestingly, the deformed drop due to gravity against the surface tension dries faster than that sitting on a horizontal substrate. For tilt angles below $\pi/2$, this enhancement of evaporation is monotonically dependent on the increase of the tilt angle. An apparent heat resistance is proposed to explain this faster evaporation behavior. In addition, the singularity problem of the evaporation flux at the contact line was physically removed with {the} concept of the thin film but with a finite thickness, regarded as an intrinsic heat resistance. 

In the present model, a Knudsen-like gas kinetics approach is employed for the evaporation flux  (Eq. \ref{EFlux}); this effect is generally extremely rapid, and the liquid molecules leave the vapour-liquid interface at the speed of thermal velocity, on the order of several hundred meters per second. The energy needed for evaporation is compensated by heat diffusion. Thanks to the short time scale of Knudsen effect, the dominant factor is therefore the `slow' heat diffusion, called `heat diffusion-limit regime'.  Therefore, our analysis focuses on the liquid side: the loss of energy in the form of the evaporation latent heat is reasonably balanced by the heat diffusion flux across the liquid layer inside the drop under the temperature gradient. With this energy (heat) balance, the local evaporation is thus calculated with the proposed herein the heat resistance concept. This is the general physical picture here.

In the end, it should be noted that the proposed model is not only valid for a given static profile, but it is fully capable for a time evolution discussion as well. At the initial time, the known parameters are the contact length $d_{SL}$, the volume $V$ of the liquid, and physical parameters of the liquid, with which we are able to determine the profile of the liquid - vapor interface and further the local evaporation flux $J$ along it. Then, in a well selected time interval $d t$, we are able to estimate the evaporation quantity  $dV=\int J(x,t) dx dt$, where $J(x,t)$ is known. After the time period $dt$, the liquid quantity becomes $V-dV$; together with $d_{SL}$ and the physical parameters, we are able to compute the new profile and the local evaporation flux along it, and then determine the evaporation quantity $dV$ in the second time interval, and so forth. In this way, we are able to discuss the time evolution dynamics of the drop evaporation, which is actually in the same spirit as numerically solving a fourth order differential equation (Eq. \ref{Ht}). Besides, the technique we developed here is suitable for the cases where the symmetry boundary conditions are not satisfied. We choose not to explicitly discuss the time evolution of the drop evaporation, because it is not the central message we would like to deliver and it is just a linear repetition of the proposed procedure, in order not to distract the readers’ attention from the local evaporation flux of the deformed drops and the associated singularity problem. 

\color{black}
\section*{Appendix A: Finding profiles of deformed drops due to gravity }
\setcounter{equation}{0}
\renewcommand{\theequation}{A\arabic{equation}}
\begin{figure*}[t]
  \includegraphics[width=0.8\textwidth]{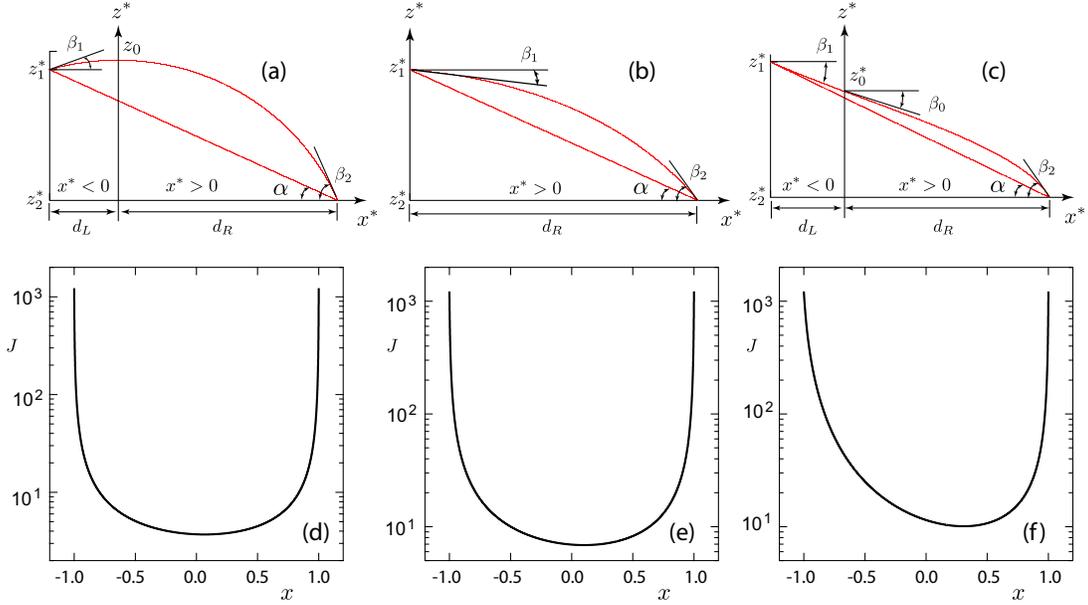}
\caption{ Drop profiles and evaporation fluxes of the three scenarios on a tilt substrate considered by Lv and Shi \cite{Lv2018PRE}. Panels (a), (b) and (c) are respectively for $\theta_r > \alpha$, $\theta_r \leq \alpha$, and $\theta_r \leq \alpha$ (the profile includes a concave and convex part), with $\theta_r$ the receding contact angle; the corresponding evaporation fluxes are presented in panels (d), (e) and (f).}
\label{3Scenarios}
\end{figure*}
In the appendix, we present the way to find the drop profiles in this paper. 

The principle employed herein has been detailedly discussed by Lv and Shi\cite{Lv2018PRE}. The authors considered three different scenarios for drops on a tilt substrate. The profiles for all the three cases can computed using Young-Laplace equation (Eq. \ref{YL}), by applying the corresponding boundary conditions, and we summarise briefly in the following the expressions for profiles, including the trivial case for the horizontal substrate.

For a drop siting on a horizontal substrate under gravity (Fig. \ref{DropSchematics}a), we can describe the drop profile by the following equations
\begin{eqnarray}
x^* = \pm \frac{\sqrt{2}l_c}{2} \int_0^{\eta} \frac{\cos{\xi}}{\sqrt{A - \cos{\xi}}} {\rm d}{\xi}, \quad \eta \in [0,\theta],
\label{Eq_1}
\end{eqnarray}
\begin{eqnarray}
z^* = - \sqrt{2} l_c \sqrt{A-\cos{\eta}}, \quad \eta \in [0,\theta].
\label{Eq_2}
\end{eqnarray}

\begin{figure}
  \includegraphics[width=0.5\textwidth]{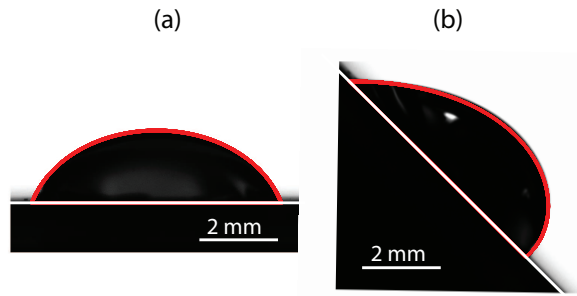}
\caption{Experimental verification of liquid drops profiles under gravity. (a)Shape of a water drop in experiments on the horizontal substrate ($\alpha = 0^{\circ}$), and (b) tilted substrate ($\alpha = 43^{\circ}$), with the computed profiles in red lines.}
\label{DropProfile}
\end{figure}

As for the drops on tilt substrates, expressions for the above-mentioned three scenarios are lists in the following. (1) Receding contact angle is above the tilt angle $\theta_r > \alpha$ (Fig. \ref{3Scenarios}a): when $x^* \le 0$, we have
\begin{eqnarray}
x^* = - \frac{\sqrt{2}l_c}{2} \int_0^{\eta} \frac{\cos{\xi}}{\sqrt{A - \cos{\xi}}} {\rm d}{\xi}, \quad \eta \in [0,\beta_1],
\label{Eq_3}
\end{eqnarray}
\begin{eqnarray}
z^* = - \sqrt{2} l_c \sqrt{A-\cos{\eta}}, \quad \eta \in [0,\beta_1];
\label{Eq_4}
\end{eqnarray}
when $x^* \ge 0$, we have
\begin{eqnarray}
x^* = \frac{\sqrt{2}l_c}{2} \int_0^{\eta} \frac{\cos{\xi}}{\sqrt{A - \cos{\xi}}} {\rm d}{\xi}, \quad \eta \in [0,\beta_2],
\label{Eq_5}
\end{eqnarray}
\begin{eqnarray}
z^* = - \sqrt{2} l_c \sqrt{A-\cos{\eta}}, \quad \eta \in [0,\beta_2].
\label{Eq_6}
\end{eqnarray}
(2) Receding contact angle is below the tilt angle $\theta_r \leq \alpha$ (Fig. \ref{3Scenarios}b). In this case, we have only $x^*>0$, and profile is given by
\begin{eqnarray}
x^* = \frac{\sqrt{2}l_c}{2} \int_0^{\eta} \frac{\cos{\xi}}{\sqrt{A - \cos{\xi}}} {\rm d}{\xi}, \quad \eta \in [\beta_1,\beta_2],
\label{Eq_7}
\end{eqnarray}
\begin{eqnarray}
z^* = - \sqrt{2} l_c \sqrt{A-\cos{\eta}}, \quad \eta \in [\beta_1,\beta_2].
\label{Eq_8}
\end{eqnarray}
(3) Receding contact angle is below the tilt angle $\theta_r \leq \alpha$ and the profile consists of a concave and a convex part (Fig. \ref{3Scenarios}c). In this case,  $\beta_1 < 0$, and $-\beta_1 > \beta_0 >0$ \cite{Lv2018PRE}. When $x^* \le 0$, we have
\begin{eqnarray}
x^* = - \frac{\sqrt{2}l_c}{2} \int_{\beta_0}^{\eta} \frac{\cos{\xi}}{\sqrt{A - \cos{\xi}}} {\rm d}{\xi}, \quad \eta \in [\beta_0,-\beta_1],
\label{Eq_9}
\end{eqnarray}
\begin{eqnarray}
z^* = \sqrt{2} l_c \sqrt{A-\cos{\eta}}, \quad \eta \in [\beta_0,-\beta_1];
\label{Eq_10}
\end{eqnarray}

when $x^* \ge 0$, we have
\begin{eqnarray}
x^* = \frac{\sqrt{2}l_c}{2} \int_{\beta_0}^{\eta} \frac{\cos{\xi}}{\sqrt{A - \cos{\xi}}} {\rm d}{\xi}, \quad \eta \in [\beta_0,\beta_2],
\label{Eq_11}
\end{eqnarray}
\begin{eqnarray}
z^* = - \sqrt{2} l_c\sqrt{A-\cos{\eta}}, \quad \eta \in [\beta_0,\beta_2].
\label{Eq_12}
\end{eqnarray}
In the above equations, there is an unknown parameter $A$. Actually, for a given system, $A$ can be determined \cite{Lv2018PRE}, and then we can plot the profile of the drop. In other words, $A$ is an {\it a priori} given parameter. With the obtained profiles, we are able to compute easily $d_L$ and $d_R$. Then a  translation and a rotation of the coordinates are done to convert $(x^*, z^*)$ in Fig. \ref{3Scenarios}a, b and c to $(x, z)$ in Fig. \ref{DropSchematics}, which are given by the following equations
\begin{eqnarray}
x = (x^* - \frac{d_R-d_L}{2}) \cos \alpha - (z^* - \frac{z^*_1 + z^*_2}{2})\sin \alpha,\quad
\label{Eq_13}\\
z = (x^* - \frac{d_R-d_L}{2}) \sin \alpha + (z^* - \frac{z^*_1+ z^*_2}{2})\cos \alpha.\quad
\label{Eq_14}
\end{eqnarray}
To verify these equations, we append the computed profiles to the corresponding experimental images. Two examples are presented in Fig. \ref{DropProfile}. In the trivial horizontal case, the measurements from experiments shows a smooth lens-shape of the drop, which has the nice overlap of the vapour-liquid interface with the profile calculated using the same parameters of contact diameter, surface tension of water, gravity constant in the experiment (Fig. \ref{DropProfile}a). The same consistence has been observed for the drop on the slope with a tilt angle of 43$^\circ$ between the experimental observation and the calculation(Fig. \ref{DropProfile}b).

We have encountered all above-mentioned scenarios when discussing the evaporation, as shown in Fig. \ref{3Scenarios}a, b and c. The difference lies only in the ways finding the profiles, as summarised in this { \it Appendix}. As for the evaporation flux, there is no difference in physics showing, one just needs to follow Eq. \ref{JJ}. Accordingly, the evaporation fluxes for the three scenarios are respectively presented in Fig. \ref{3Scenarios}d, e and f, under the same condition as in Section II.
\color{black}
\section*{Acknowledgements}
P. J. is supported by Harbin Institute of Technology (Grant No. HA45001103 and HA11409052). G. J. and C. L. thank the supporting grants NSFC 11774287 and {11872227}. H.Y. acknowledges National Science Foundation of EDS (203010036). 

\section*{Data availability}
The data that support the findings of this study are available from the corresponding author upon reasonable request.

\bibliography{Evaporation_PRF}
\end{document}